\documentclass[preprint,aps,floatfix]{revtex4-1}

\usepackage[T1]{fontenc}       

\usepackage{ae,aecompl}
\usepackage{graphicx}
\usepackage{epstopdf}
\usepackage{color}
\usepackage{amsmath,amssymb}
\usepackage{hyperref}
\usepackage{amsfonts}

\newcommand{\comment}[1]{}

\begin{document}

\title{From Hindered to Promoted Settling in Dispersions of Attractive Colloids: Simulation, Modeling, and Application to Macromolecular Characterization}

\author{Andrew M. Fiore}
\author{Gang Wang}
\author{James W. Swan}
\email{jswan@mit.edu}
\affiliation{Department of Chemical Engineering, Massachusetts Institute of Technology, Cambridge, MA 02139, USA}


\begin{abstract}
The settling of colloidal particles with short-ranged attractions is investigated via highly resolved immersed boundary simulations.  At modest volume fractions, we show that inter-colloid attractions lead to clustering that reduces the hinderance to settling imposed by fluid back flow.  For sufficient attraction strength, increasing the particle concentration grows the particle clusters, which further increases the mean settling rate in a physical mode termed \emph{promoted} settling.  The immersed boundary simulations are compared to recent experimental measurements of the settling rate in nanoparticle dispersions for which particles are driven to aggregate by short-ranged depletion attractions.  The simulations are able to quantitatively reproduce the experimental results. We show that a simple, empirical model for the settling rate of adhesive hard sphere dispersions can be derived from a combination of the experimental and computational data as well as analytical results valid in certain asymptotic limits of the concentration and attraction strength.  This model naturally extends the Richardson-Zaki formalism used to describe hindered settling of hard, repulsive spheres.  Experimental measurements of the collective diffusion coefficient in concentrated solutions of globular proteins are used to illustrate inference of effective interaction parameters for sticky, globular macromolecules using this new empirical model.  Finally, application of the simulation methods and empirical model to other colloidal systems are discussed.
\end{abstract}

\maketitle


\section{Introduction}
Settling of dispersed colloidal particles is central to the processing and analysis of a wide range of industrial and scientific materials. Centrifugation is used as both a processing and analytical tool in laboratory and commercial settings\cite{auc}, and gravity-driven particle motion controls the shelf life of many consumer and food products\cite{shelflife}. In the context of environmental science, sedimentation of suspended particles plays a crucial role in the engineering of pollution remediation strategies\cite{pollution} as well as natural processes of erosion and deposition\cite{erosion}. Additionally, there is a linear relationship between the sedimentation coefficient and the collective diffusivity in colloidal dispersions. Thus, important industrial separation processes such as ultrafiltration applied to purification of biologically derived macromolecules, whose design requires accurate models of the collective diffusivity, depend on knowledge of the sedimentation coefficient in concentrated dispersions\cite{ultrafiltration}. In a vast majority of these materials and systems, the colloidal particles are attractive. However, most fundamental studies of settling in colloidal dispersions concern hard or repulsive particles. In the present work, we apply the immersed boundary method to study the settling rate in concentrated dispersions of spherical colloids with short-ranged attractions.

The settling dynamics of concentrated colloidal dispersions are controlled by the hydrodynamic interactions induced as moving particles displace the surrounding fluid. The flows generated by settling particles decay to leading order as the inverse of the distance from the particle.  The entrainment of other particles by these disturbance flows is a critical factor in settling dispersions. The combined flow produced by all the settling particles appears to diverge with the physical size (extent) of the dispersion because of the long-ranged nature of these hydrodynamic interactions\cite{guazzelli}. However, mass conservation at the system boundaries produces a back-flow driven by a pressure gradient balancing the buoyant weight of the particles. When the back-flow is combined with the induced flows, an intensive sedimentation rate, one that is independent of the system size, results. The interplay of entrainment and back-flow is a subtle one that makes the sedimentation a sensitive function of the microstructure of the colloidal dispersion. In dispersions of attractive colloids, which tend to form clusters, entrainment and back-flow act synergistically so that attractions always accelerate the settling process. The reasons for this are straightforward: clusters in isolation naturally sediment more quickly than individual particles, and the widened spaces between clustering particles lead to a weaker resistance due to the back-flow. Quantifying these effects represents a major challenge in quantitative colloid science. In the present work, we consider a limit in which the P\'eclet number for the settling particles is vanishingly small. In this limit, the rate of particle diffusion far exceeds the rate of sedimentation and the microstructure of the colloidal dispersion is dictated by thermodynamic equilibrium.

Of the limited studies on the sedimentation of attractive suspensions, the most important result, a dilute-limit approximation for the mean settling rate $U$ of a suspension of spherical particles, accurate to first order in the particle volume fraction $\phi$, was derived by Batchelor more than fifty years ago\cite{batchelor1,batchelor2}:
\begin{equation}
	\frac{U}{U_{0}} = 1 - 6.55 \, \phi + 3.52 \, \left( 1 - B_{2}^{\ast} \right) \, \phi + \mathcal{O}\left(\phi^{2}\right), \label{eqn:batchelor}
\end{equation}
where $U_{0}$ is the settling speed of a single isolated particle and $B_{2}^{\ast}=B_{2}/B_{2}^{HS}$, where $B_{2}$ is the second virial coefficient of the particles and $B_{2}^{HS}$ is the second virial coefficient of a reference hard sphere dispersion. For dilute hard spheres, entrainment and back-flow combine to reduce the relative settling rate of particles by an amount of $6.55\,\phi$.  Attractive interactions cause $ B_{2}^* $ to drop from unity (hard spheres) to negative values. This has the effect of increasing the settling speed relative to hard spheres.  For sufficiently strong attractions ($B_{2}^{\ast} \leq -0.861$), Batchelor's theory predicts that a dispersion can settle faster than an isolated particle. \citeauthor{monchojorda1} used a series of stochastic rotation dynamics (SRD) simulations to test Batchelor's dilute limit prediction as well as to study sedimentation in concentrated attractive dispersions. They observed that particles with attractive interactions sediment more quickly than hard spheres at all volume fractions and that, within a specific range of volume fractions, sufficiently strong attractions produce a settling rate that is larger than even the isolated particle value. For dispersions that exhibited such promoted settling, the sedimentation rate exhibited a non-monotonic dependence on particle volume fraction. At low volume fractions, the rate increases above the single particle value. Beyond a critical volume fraction, the settling rate decreased as expected of conventional hindered settling. These simulations and the resulting data possess a number of physical and quantitative issues, including: being performed at finite but small P\'eclet and Reynolds numbers, being limited in the number of particles represented in the periodic simulation box, and having made no systematic finite system size corrections to the measured sedimentation rates. Inspired by these simulations, \citeauthor{lattuada} conducted an experimental study designed to corroborate the simulation results\cite{lattuada}. They found a qualitative agreement between the experiments and simulations. However, they observed that the SRD simulations systematically under-predict the settling rate of attractive suspensions measured experimentally and also under-predict the range of volume fractions over which promoted settling is observed. These studies are the first steps towards understanding the settling of concentrated solutions of attractive particles, but much remains unexplored. For example, a predictive microstructural model for the settling rate would be very useful in applications involving attractive colloidal dispersions. 

In the present work we use immersed boundary simulations in the limits of zero P\'eclet and Reynolds numbers to systematically investigate the effect of attraction strength and particle volume fraction on the mean settling rate. We show that these simulations, which account properly for finite system size effects reproduce experimental measurements quantitatively.  Furthermore, we develop an empirical model for the settling rate based on a measure of microstructure in attractive dispersions that quantitatively matches experimental data and our own simulation data at all the volume fractions studied. 

\section{Simulation Methodologies}
We compute the settling rate of attractive colloidal dispersions using an immersed boundary method referred to colloquially as the composite bead\cite{gang} or rigid multi-blob\cite{multiblob} approach. The details of the computational method are described elsewhere\cite{gang}. Here we give a brief review. In this approach, the surface of each colloidal particle is approximated by a collection of beads that interact hydrodynamically.  Each bead generates a regularized Stokeslet flow and is entrained by flows induced by other particles in a reciprocal fashion\cite{pse,pse-stresslet}. In our formulation, the Rotne-Prager-Yamakawa tensor\cite{rpy} that linearly relates the force exerted on one bead to the entrained velocity of another bead is used to represent these flows. Additionally, the beads tessellating the surface of each particle are constrained to move as a rigid body. A set of Lagrange multipliers, forces that ensure the constraints are satisfied, are introduced and a system of linear equations is solved to determine the rigid body motions of the colloidal particles and the Lagrange multipliers in response to any set of imposed external forces. In the present work, the spectral Ewald method\cite{lindbo-tornberg} is used to evaluate the product of the Rotne-Prager-Yamakawa tensor for all the beads with an arbitrary set of forces acting on the beads. The linear equations governing the rigid body motions and the Lagrange multipliers are solved using the GMRES methodology with constraint preconditioning\cite{keller}. In summary, this approach can be used to evaluate the transport properties of macromolecular and colloidal solutions of arbitrarily shaped particles in $ O( N ) $ time where $ N $ is the total number of beads used to discretize the surface of all the colloids.

\begin{figure}
	\includegraphics[width=0.49\textwidth]{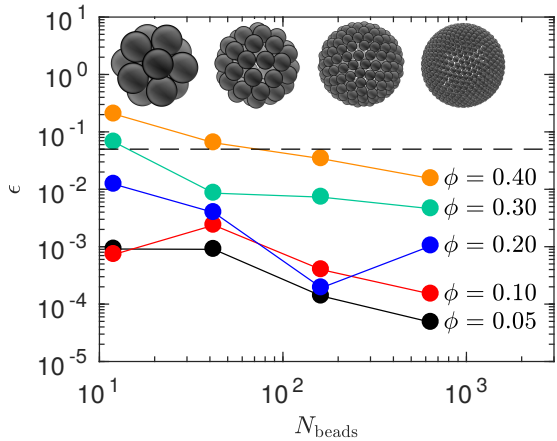}
	\caption{Relative error in the mean sedimentation rate of a periodic simple cubic lattice of spherical particles computed using the icosphere approximation of the spherical surface compared to a high resolution approximation ($N_{\rm bead}=2562$) as a function of the number of vertices of the polyhedron, $N_{\rm bead}$, at different volume fractions, $\phi$. We find that $N_{\rm bead}=2562$ is sufficient to exactly reproduce the published data of \citeauthor{zickhomsy}\cite{zickhomsy}. The polyhedra used in the calculations are shown above the associated points. The black dashed line denotes 5\% relative error. }
	\label{fig:convergence}
\end{figure}

For the spherical colloids studied here, the beads are made to sit on the vertices of an icosphere formed by subdividing the faces of a Goldberg polyhedron (see Figure \ref{fig:convergence}). We have shown how such a discretization can be used to replicate the hydrodynamic interactions among a pair of spherical particles in a previous publication\cite{gang}. Convergence of the calculation with respect to the number of beads is slow for relative motion between particles because of the effects of hydrodynamic lubrication. However, convergence is fast for collective modes of motion. 
Figure \ref{fig:convergence} also depicts the convergence of the sedimentation rate of a cubic array of spherical colloids as a function of the number of beads tessellating the surface. \citeauthor{zickhomsy} performed high accuracy calculations of lattice sedimentation and published their results with three digits of accuracy\cite{zickhomsy}. We find that 2562 beads per colloid is sufficient to exactly reproduce the published results, and therefore use the solution with 2562 as the reference state for the convergence study. With just 162 beads per colloid, the classic results of \citeauthor{zickhomsy} are recovered to within 5\% for the full range of volume fractions studied. Therefore, in the present calculations we use 162 beads per colloid. This level of discretization is chosen to accommodate a balance between the number of colloids required to accurately model the microstructure of an attractive dispersion, the number of realizations of that microstructure needed to compute statistically meaningful averages, and the total computation time of the simulation. It is known that number density fluctuations are large in attractive colloidal dispersions\comment{\cite{poon}}. Additionally, the magnitude of velocity fluctuations among sedimenting colloids can be of the same order of magnitude as the mean sedimentation velocity\cite{segre}. Thus, $ 1000 $ colloids are modeled and averages of the settling velocity are constructed from $ 1000$ independent snapshots of the dispersion microstructure. Throughout this work, fluctuations in the settling rate among different configurations are used to quantify a standard error at the 95\% confidence level.

The experiments of \citeauthor{lattuada} control inter-particle attraction by adding surfactant micelles to solution in order to induce depletion\cite{lattuada}. Therefore, we model the interaction potential, $ U(r)$ with a short-ranged attraction described by the Asakura-Oosawa depletion potential\citep{ao_potential} plus a hard core repulsion at inter-particle contact, $ r = 2a $:
\begin{equation}
	V( r ) = \left\{ \begin{array}{cc} \infty & r < 2 a \\ 
					-\varepsilon \, \frac{\left( a + \delta \right)^{3}}{\delta^{2} \, \left( 1.5a + \delta \right)} \, \left( 1 - \frac{3}{4}\frac{r}{a+\delta} + \frac{1}{16}\left(\frac{r}{a+\delta}\right)^{3} \right) & r \ge 2 a  \end{array} \right.  ,
\end{equation}
where $ \varepsilon $ is the strength of the attraction at contact and $\delta$ is the range of the attraction. The Asakura-Oosawa potential was validated in these experiments through measurements of the osmotic compressibility in the colloidal dispersion. The range of the depletion potential is chosen to match the experiments: $\delta/a = 0.028$. The depletion potential at contact, $\varepsilon$, is varied and can be chosen so that simulated dispersions match the experimentally measured second virial coefficient, $B_{2}^{\ast}$. Consistent with experiments, 5\% dispersity in particle size is introduced to suppress crystallization in the simulations. 

The experimental value of the P\'{e}clet number is small, $\mathrm{Pe} = 4\pi \Delta\rho g a^{4} / 3 k_{B}T \approx 10^{-3}$, where $\Delta \rho$ is the density difference between the immersed colloids and the fluid, $g$ is the gravitational constant, and $k_{B}$ is Boltzmann's constant, so the distribution of particle positions during settling is very close to the equilibrium distribution. Therefore, simple, freely draining Brownian dynamics simulations of single beads are used to generate representative configurations of the particles from which the mean settling rate can be computed. For each combination of $\phi$ and $ \varepsilon $ (or $B_{2}^{\ast}$), an initial hard sphere configuration of spherical colloids is allowed to relax for 1000 bare diffusion times ($ 6 \pi \eta a^ 3 / k_B T $). Then, snapshots of the particle configuration  are taken every 10 bare diffusion times until 1000 snapshots have been accumulated. For each snapshot, a composite-bead representation of all the particles is constructed from 162 bead icospheres having randomly assigned orientations. The mean settling rate for each configuration is determined by applying a uniform force to each bead and computing the mean of the resulting rigid body translational velocities of the colloids. The computed mean velocity is sensitive to the periodic boundary conditions in the simulations, and the infinite system size limit $U^{\infty}$ is found by
\begin{equation}
	U^{\infty} = U + 1.7601 \, S(0) \, \left( \frac{\eta_{s}}{\eta\left(\phi\right)} \right) \, \left( \frac{\phi}{N} \right)^{1/3}, \label{eqn:size_correction}
\end{equation}
which is the finite size correction given by \citeauthor{laddweitz}\cite{laddweitz}, where $S(0)$ is the value of the static structure factor at zero wavenumber, $\eta_{s}$ is the solvent viscosity, and $\eta\left(\phi\right)$ is the high-frequency viscosity, which we compute for each configuration. It should be noted that, although we use the depletion potential to model the particle attractions, with this procedure any sufficiently short-ranged potential will produce equivalent equilibrium structures\cite{noro}. The Noro-Frankel rule of corresponding states suggests that samples of these equilibrium structures can be drawn from a mapping onto the Baxter adhesive hard-sphere interaction potential. In section \ref{sec:ahs_map} we discuss this mapping explicitly. Therefore, the results presented in this manuscript should be generic for any colloidal dispersion aggregating due to short-ranged attractions. 

\section{Results and Discussion}

	\subsection{Comparison of Simulations and Experiments}
\begin{figure}
		\includegraphics[width=0.49\textwidth]{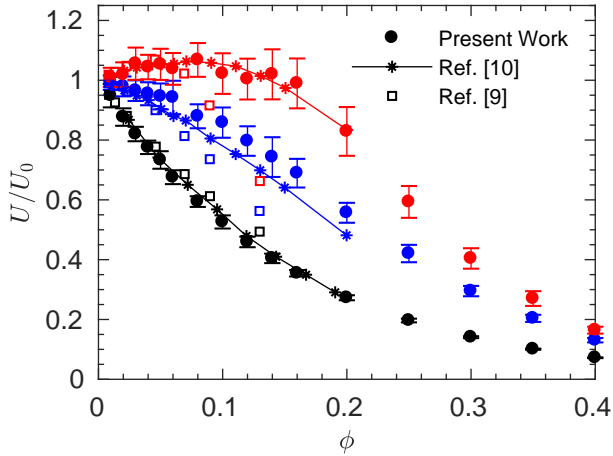}
		\caption{Normalized settling velocity as a function of particle volume fraction for $B_{2}^{\ast} = -1.08$ (red), $B_{2}^{\ast} = -0.27$ (blue), and $B_{2}^{\ast} = 1$ (black) from the experimental data of Ref. \cite{lattuada} (stars with lines) and computed in the present work (filled circles) and the SRD simulations of Ref. \cite{monchojorda1} as interpolated in Ref. \cite{lattuada} (open squares). Error bars represent the standard error of the observed measurements.}
		\label{fig:velocity_comparison}
	\end{figure}

	There is limited experimental data available for the mean settling rate of colloids as a function of concentration for different well-characterized attraction strengths. \citeauthor{lattuada} provide data at three different attraction strengths, characterized by the reduced second virial coefficients: $B_{2}^{\ast} = 1.0, -0.27, -1.08$\cite{lattuada}. These values correspond to hard sphere interactions (no attraction), a modest attraction ($\varepsilon = 4.84 \, k_B T $), and a strong attraction  ($\varepsilon = 5.18 \, k_B T $) that is near the boundary for liquid-liquid phase separation, respectively. The experiments were performed with particle volume fractions, $\phi$, as large as $\phi=0.20$. We compute the mean settling rate for each of the experimental values of $B_{2}^{\ast}$ for $0.01 \leq \phi \leq 0.40$.  Figure \ref{fig:velocity_comparison} shows our simulation results, the experimental data of \citeauthor{lattuada}, and the simulation data of \citeauthor{monchojorda1}. Our results are in quantitative agreement with the experimental data for all values of $B_{2}^{\ast}$ and $\phi$ for which a comparison is available. The SRD simulation results of \citeauthor{monchojorda1} display only qualitative agreement with our simulations and the experimental data, and predict a much sharper decay of $U$ with $\phi$ than is observed experimentally for the two attractive dispersions studied. 
	The agreement between experiments and our simulations affirms our claim that the proposed simulation methodology is appropriate to perform a systematic study the sedimentation rate as a function of particle volume fraction and attraction strength. 

It is a challenge to assess the discrepancy between the SRD simulations and our own. A number of factors may be at play. One central problem is a failure of \citeauthor{monchojorda1} to account for finite system size effects in the calculation of the settling rate. As in Equation \eqref{eqn:size_correction}, in periodic geometries transport properties like settling rate and diffusivity are depressed by an amount proportional to $ a V^{-1/3} $ where $ V $ is the volume of the simulation cell. For simulations with a fixed number of particles, increasing the volume fraction acts to decrease $ V $, which further depresses the computed values of transport properties. However, to change volume fraction, \citeauthor{monchojorda1} fix the simulation volume and increase $ N $. So a systematic depression of the settling rate due to finite system size effects alone is not enough to explain their data. A further issue that may explain the discrepancy with their simulations is the finite P\'eclet number used in that study. The authors estimate that $ \mathrm{Pe} \approx 2.5 $ in their \emph{dynamic} simulations, which means that deviations from equilibrium distribution for the particle configuration are inevitable. The effect of $ \mathrm{Pe} $ on the mean settling rate of dispersions is also not well studied. However, for hard-spheres in the dilute limit and at large Peclet numbers, Cichocki and Sadlej \cite{cichocki} have shown that a different expression for the sedimentation rate is expected: $ U/U_0 = 1 - 3.87 \phi $. The settling rate is a sensitive function of the structure, and the settling process modeled dynamically by the SRD simulations may have exhibited changes from the equilibrium structure and even structural anisotropy promoted by the hydrodynamic interactions among the particles. In the present study, the Peclet number is set asymptotically to zero and only the equilibrium structure is used for the calculation of the transport properties.

	\subsection{Microstructural Characterization and Corresponding States}
	\label{sec:ahs_map}
	As is common for colloids with short-ranged attractions, \citeauthor{lattuada} characterized the attraction strength in their dispersion using the Baxter temperature, $\tau$. The Baxter temperature is a parameter that arises in the adhesive hard sphere (AHS) model of attractive particles, in which the inter-particle potential has vanishing width and infinite depth, but is characterized by a single parameter, $ \tau $, reflecting the propensity of particles to adhere to one another\cite{baxter}. This parameter is a measure of the effective strength of attraction in the suspension, with smaller values corresponding to stronger attractions. $B_{2}^{\ast}$ for the AHS model can be related to the Baxtern temperature, $\tau$, as $B_{2,{\rm AHS}}^{\ast} = 1 - 1/\left(4\tau\right)$. In the approach taken by \citeauthor{lattuada}, $\tau$ is computed by matching the measured $B_{2}^{\ast}$ to the analytical expression for $B_{2}^{\ast}$ of AHS model, so that:
\begin{equation}
	\tau = \frac{1}{4} \left( 1 - B_{2}^{\ast} \right)^{-1}. \label{eqn:ahs}
\end{equation}  
Assigning a value to $ \tau $ in this fashion provides a way of determining the dilute limit thermodynamic and transport properties of dispersions with short-ranged attractions in a fashion that is agnostic to the details of the interaction potential. However, values of $\tau$ computed by \eqref{eqn:ahs} are valid only in the dilute limit. As the dispersion concentration increases, $\tau$ changes even for fixed $B_{2}^{\ast}$, and as a result more detailed models are required to infer $\tau$ for dispersions of arbitrary concentration. 
		
	In the original development of the AHS model, Baxter produced an analytical expression for the Percus-Yevick (PY) approximation to the pair distribution function $g(r)$ of sticky particles\cite{baxter,percus-yevick}. The AHS model results in an expression for $g(r)$ that is completely parameterized by the particle size $a$, $\phi$, and the effective ``stickiness" parameter -- the Baxter temperature -- $\tau$. An analytical expression for the static structure factor, $S(q)$, of an AHS suspension can be directly computed from Baxter's result\cite{regnaut2}. Applying the Noro-Frenkel rule of corresponding states, which states that the thermodynamic properties, including dispersion microstructure, of colloids with short-ranged attractions can all be mapped back onto the adhesive hard-sphere model, the stickiness parameter, $\tau$, for a concentrated dispersion of colloids with short-range attractions can be inferred from calculations of either $g(r)$ or $S(q)$ (or any other state function) for that dispersion. For any of these functions, the calculated values can be fit by analytical expressions from the adhesive hard sphere model at the same particle concentration to determine a corresponding value for $ \tau $. In the dilute limit, using the osmotic pressure as the state function will recover \eqref{eqn:ahs}. For more concentrated dispersions, $S(q)$ (or $g(r)$) is required to determine the particular value of $ \tau $ that characterizes the thermodynamic state of the dispersion in the corresponding adhesive hard-sphere model. For this reason, throughout this manuscript $\tau$ is an \textit{a priori} unspecified parameter which is inferred from simulation data. 

\begin{figure}
		\includegraphics{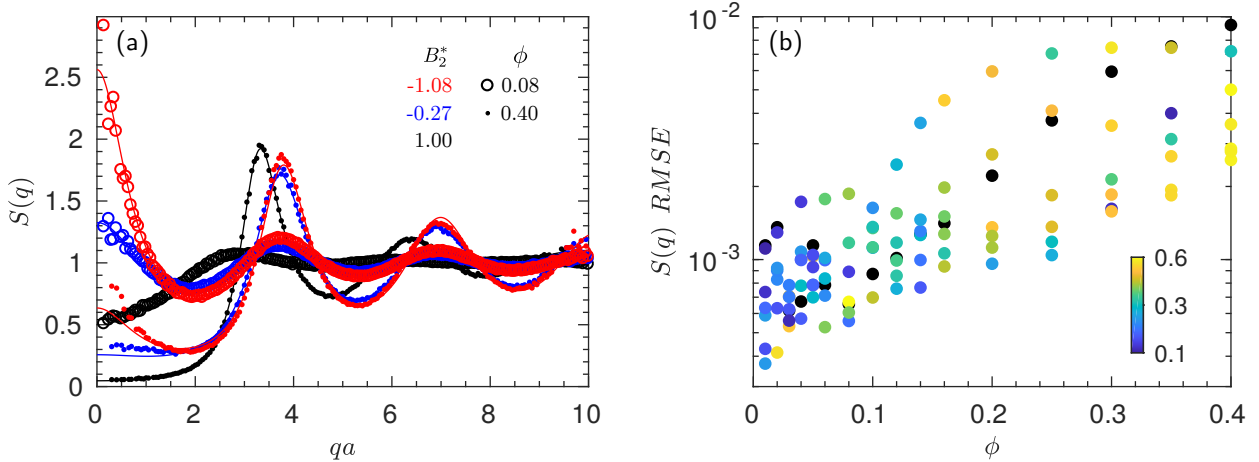}
		\caption{ (a) Static structure factor computed for the three attraction strengths reported in Figure \ref{fig:velocity_comparison}, $B_{2}^{\ast} = -1.08$ (red), $B_{2}^{\ast} = -0.27$ (blue), and $B_{2}^{\ast} = 1$ (black), at $\phi=0.08$ (open circles) and $\phi=0.40$ (dots). The lines are fits by the AHS solution to the Percus-Yevick approximation for $S(q)$. The $\tau$ values computed for the fit are 10.53, 0.16, and 0.11 for the black, blue, and red data sets, respectively at $\phi = 0.08$ and 6.53, 0.21, 0.12 for the respective data sets at $\phi = 0.40$. (b) Root-mean-squared error of the AHS Percus-Yevick solution to the simulation data colored by the inferred $\tau$ value, with black corresponding to hard spheres.}
		\label{fig:sofq}
	\end{figure}

	Figure \ref{fig:sofq} shows the values of $S(q)$ computed at the three attraction strengths from Figure \ref{fig:sofq} at $\phi=0.08$ and $\phi=0.40$, compared to the PY approximation for adhesive hard spheres using a value of $\tau$ inferred by minimizing the root mean-squared error (RMSE) of the PY model with respect to the computed data. Excellent agreement is observed between the model and the data for the six examples shown. Similar agreement in of the PY model with the fit structure factor is associated with other particle concentrations and attraction strengths.  
	Figure \ref{fig:sofq}b shows the relative error in the computed static structure factor with respect to the Percus-Yevick approximation at the best fit value of $ \tau $ as a function of the colloid volume fraction and the second virial coefficient for all the simulated dispersions.  Errors smaller than 1\% are typical, indicating that the fitting procedure is adequate for characterizing the suspension microstructure in terms of the generalized stickiness parameter. This fitting procedure can be generalized to experimental data with sticky macromolecules of known number density, $ n $ as discussed in section \ref{sec:expt}.  In this way, measurements of $S(q)$, allows mapping onto the AHS model without \textit{a priori} knowledge of either $B_{2}^{\ast}$.  The pair $ \tau $ and the volume fraction serve as unique descriptors of the attractive suspension microstructure and as we will show, the value of this pair also fixes the mean settling rate. 	  
	 
	\subsection{Settling rate as a function of $ \phi $ and $ \tau $}
\begin{figure}
		\includegraphics[width=0.99\textwidth]{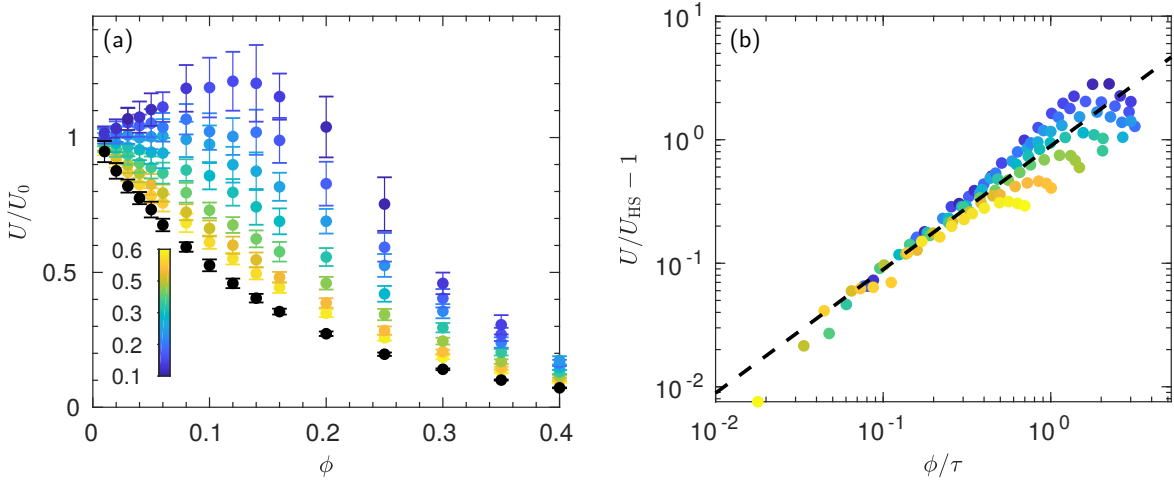}
		\caption{(a) Normalized settling velocity as a function of particle volume fraction for $-1.48 \leq B_{2}^{\ast} \leq 1.00$ colored by the value of the stickiness parameter $\tau$ inferred from the static structure factor (black is hard sphere data). (b) Relative difference between the settling rate of attractive and hard sphere dispersions for each condition presented in panel (a). The dashed black line denotes the Batchelor result in the limit $ \phi \rightarrow 0 $: $0.89 \phi/\tau$. }
		\label{fig:velocity_allsims}
	\end{figure}

	We performed simulations systematically varying $B_{2}^{\ast}$ between $-1.48$ and $1.00$ and $\phi$ from $0.01$ to $0.40$ for a dispersion of spherical colloids interacting via a depletion potential with $ \delta / a = 0.028 $. Each set of parameters is characterized by the $\tau$ value inferred from $S(q)$. Shown in Figure \ref{fig:velocity_allsims}a is the mean settling rate as a function of volume fraction and inferred Baxter temperature normalized by the single particle value, $ U/U_{0}$, where the color of the symbols corresponds to the inferred value of $ \tau $, with the results for hard spheres are depicted in black.
	
	At a fixed concentration, as $\tau$ increases, the settling velocity decreases and approaches the hard sphere limit, $ \tau \rightarrow \infty $. Furthermore, although the shape of $U/U_{0}$ is strongly dependent on $\tau$ at low $\phi$, at sufficiently large $\phi$, the settling velocity exhibits a qualitative similarity for all $\tau$. With increasing concentration, the influence of the stickiness parameter on the settling rate is diminished.  In this regime, particles fill space rather homogeneously, regardless of $ \tau $, and the sedimentation rate decays to zero much as with hard spheres.  Attractions between particles still increase the sedimentation rate slightly by introducing small heterogeneities in the microstructure that are indicated by an increase in the value of $ S(q) $ as $ q \rightarrow 0 $.  These void spaces allow back-flow to occur with less resistance, thus slightly increasing the settling rate.
	
	The simulation data show a transition in the shape of the settling curves as both $\tau$ and $\phi$ are increased. For hard spheres, $U/U_{0}$ is monotonically decreasing and concave up for all $\phi$. At modest attractions, $U/U_{0}$ is still monotonically decreasing, but is initially concave down until an inflection point around $\phi=0.2$. An inflection point is also observed for strong attractions, $ \tau \lesssim 0.2 $, but $ U/U_{0}$ is no longer monotonic.  The settling rate increases until $\phi \approx 0.10$, where it exhibits a maximum rate of promoted settling.  Then, the settling rate decays much as for hard spheres.

	The ratio of the sedimentation rate of attractive particle dispersions to a hard sphere dispersion at the same concentration is shown in Figure \ref{fig:velocity_allsims}b as a function of $\phi/\tau$.  $ \phi / \tau $ is the natural parameter that appearing in Batchelor's dilute limit result, when substituting the second virial coefficient with the AHS model\cite{batchelor2}. Batchelor's model describes the ratio well for for $\phi/\tau \lesssim 0.5$. For $\phi/\tau \gtrsim 0.5$, the data is no longer a function of $ \phi / \tau $ alone and exhibits a sharp downturn with increasing $\phi/\tau$.
	 
\begin{figure}
		\includegraphics[width=0.49\textwidth]{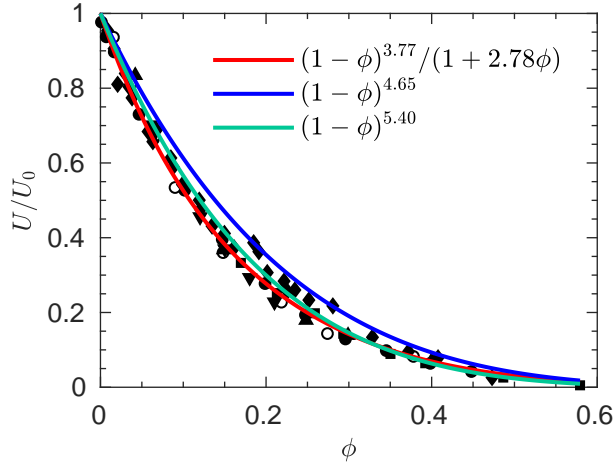}
		\caption{Settling rates for colloidal hard-spheres.  The lines represent different expressions of a Richardson-Zaki like form including \eqref{eqn:hsmodel} with best fit $ m = 3.77 $.  This particular expression fits the settling rates at high concentrations at least as well as the other expressions, but does a better job accounting for variations in the settling rate at low concentrations.  The data points are drawn from the literature: squares\cite{piazza_squares}, triangles\cite{piazza_triangles}, upside down triangles\cite{piazza_utriangles}, diamonds\cite{piazza_diamonds}, solid circles\cite{piazza_ccircles}, open circles\cite{piazza_ocircles}.}
		\label{fig:hardspheres}
	\end{figure}

	To develop an empirical model for sedimentation in attractive dispersions at arbitrary volume fractions, we start with the Richardson-Zaki correlation\cite{richardson_zaki}, which is a common model for settling in suspensions, 
	\begin{equation}
		\frac{U}{U_{0}} = \left( 1 - \phi \right)^{4.65}. \label{eqn:RZ}
	\end{equation}
	This expression is widely used in engineering applications even though it is known to deviate from Batchelor's prediction in the limit of small $\phi$. The value of the exponent in the RZ expression varies depending on the source, but the form $(1-\phi)^{m}$ is widely used, where $m$ is typically $ O(4-6) $.  For Brownian hard-spheres, a value of $ m = 5.40 $ is often used.  The correct low $\phi$ behavior can be obtained without losing accuracy at large $\phi$ by modifying the Richardson-Zaki expression to force agreement with Batchelor in the dilute limit,\cite{batchelor1}
	\begin{equation}
		\frac{U}{U_{0}} = \frac{ \left( 1 - \phi \right)^{m} } { 1 + \left( 6.55 - m \right) \phi }. \label{eqn:hsmodel}
	\end{equation}
 The value of $m$ is determined to be 3.77 by fitting Equation \eqref{eqn:hsmodel} to the hard sphere data from our simulations.  Equation \ref{eqn:hsmodel} with the best fit value for the exponent is plotted as the dashed line in Figure \ref{fig:velocity_allsims}, and the relative error with respect to the simulated data is smaller than 0.1\% for all the volume fractions studied.  The best fit power law exponent is less than the values commonly used in the Richardson-Zaki correlation, but Figure \ref{fig:hardspheres} shows that our model quantitatively describes hard sphere colloidal sedimentation data at least as well as the two other Richardson-Zaki like correlations used in the literature: $ U / U_0 = (1-\phi)^{4.65} $ and $ U / U_0 = (1-\phi)^{5.40} $.  However, \eqref{eqn:hsmodel} has the added bonus of recovering the dilute limit predictions of Batchelor precisely.
	 
	 Inspired by this approach, we notice that further modifying \eqref{eqn:RZ} with a denominator $ 1 + ( 6.55 - m ) \phi - 0.89 \phi / \tau $ would also recover Batchelor's predictions in the dilute limit for attractive dispersions.  Because the settling rate bears such qualitative similarity with hard spheres at high concentrations, we proceed with the \emph{ansatz} that the net effect of inter-particle attractions can then be accounted for by simply replacing Batchelor's: $ -0.89 \phi / \tau $ with a more general term $ -0.89 f( \phi, \tau )$.  This function, $ f( \phi, \tau ) $ must be bounded from above to avoid singularities.  It also must scale as $ \phi / \tau $ in the dilute limit and must approach zero in the hard sphere limit.  In principle this function serves to interpolate between hard sphere settling rate and the settling rate of aggregated dispersions.  The empirical expression we propose for the settling rate is,
	\begin{equation}
		\frac{U}{U_{0}} = \frac{ \left( 1 - \phi \right)^{3.77} }{1 + 2.84\,\phi - 0.89 \, f\left(\phi,\tau\right)}, \label{eqn:settling_function}
	\end{equation}
	where a form for $f\left(\phi,\tau\right)$ remains to be specified. 
	
\begin{figure}
		\includegraphics[width=0.99\textwidth]{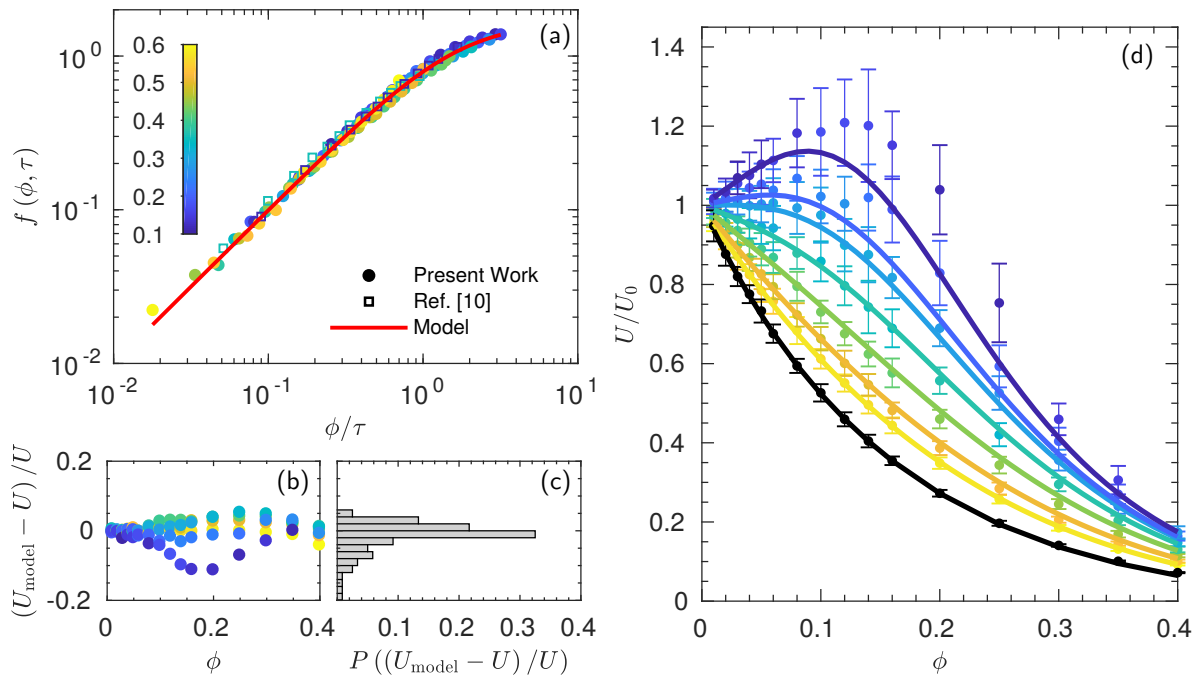}
		\caption{(a) Inter-particle interaction contribution to the hindered/promoted settling function given in Equation \eqref{eqn:settling_function} with $m=3.77$ as a function of the normalized volume fraction for a range of $\tau$ computed from sedimentation data in the present work (circles) and the experimental data of Ref. \cite{lattuada} (open squares). Hard sphere data, for which $f=0$, has been omitted. The red line is given by Equation \eqref{eqn:fitfunc}. The color of each data point corresponds to the $\tau$ value determined by fitting $S(q)$ to the AHS model. (b) Relative error in the computed model velocity $U_{\rm model}$ given by Equation \eqref{eqn:settling_function} compared to the observed velocity $U$ as a function of volume fraction. (c) Observed probability distribution for the relative error in the model prediction. (d) Settling rates predicted by the model \eqref{eqn:settling_function} for $\tau = [0.106, 0.125, 0.136, 0.165, 0.219, 0.322, 0.484, \infty]$ (lines) compared to the simulation results (Figure \ref{fig:velocity_allsims}), with the same coloring as in panel (a). Hard sphere data, for which $\tau = \infty$, is shown in black.}
		\label{fig:fitfunc}
	\end{figure}

	This hypothesized model can be tested by solving \eqref{eqn:settling_function} for $f\left(\phi,\tau\right)$, and computing this quantity from known settling rates.  Figure \ref{fig:fitfunc} shows the values of $f( \phi, \tau )$, as a function of $\phi / \tau$  with the settling rates given by our simulation data and the experimental data of \citeauthor{lattuada}.
	All the data in Figure \ref{fig:fitfunc} falls along a single master curve.  For small values of $ \phi / \tau $, $ f( \phi, \tau ) $ is linear in $ \phi / \tau $ with a slope of $ 1 $.  This is required in order to recover Batchelor's predictions in the dilute limit.  At large $ \phi / \tau $, the $ f( \phi, \tau ) $ saturates.  Note, the maximal value of $ \phi / \tau $ explored in this work is about $ 4 $, ($ \phi = 0.4$, $\tau \approx 0.1$).  For $ \tau < 0.1 $, adhesive hard spheres phase separate, and we have sought to stay within the single phase region for all the present calculations.  The function $ f( \phi, \tau ) $ can be described by a sigmoidal function of $ \phi / \tau $. There are many possible choices for this function, but a compact and convenient one is	
	\begin{equation}
		f\left(\phi,\tau\right) = \frac{\phi/\tau}{ \left[ 1 + n \, \left( \phi/\tau \right)^{p} \right]^{1/p}}, \label{eqn:fitfunc}
	\end{equation}
	where $p$ is a fitting parameter that is varied along with $n$ to minimize the sum of squared errors between the model prediction for $U$ and the data. The optimal values computed by a nonlinear least squares fit to the entire data set are $n=0.41$ and $p = 1.51$.
	
	The model accurately describes the settling rate across the wide range of $\phi$ and $\tau$ studied.  The relative error of the model with respect to the data is plotted in Figure \ref{fig:fitfunc}b.  The error is smaller than 10\% for all the cases studied.  The model does tend to slightly over-predict the settling rate at modest $\tau$ and is the least accurate for the strongest attraction studied. This can be understood by recognizing that for the most strongly attractive particles studied, $\tau \approx 0.1$, which is very close to the theoretical critical Baxter temperature, $\tau = 0.098$.  For modest $\tau$, the physical picture is that particles aggregate into transient clusters, which distribute homogeneously in the dispersion.  However, in the neighborhood of the boundary for liquid-liquid phase separation, the system exhibits large particle number density fluctuations.   This produces large fluctuations in the mean settling rate that are quantified by the standard error in Figure \ref{fig:velocity_allsims}.  In spite of this physical limitation, the results of the empirical model appear to provide good predictions of the settling rate even in this region of phase space. 

	\subsection{Inferring macromolecular interactions through application of the settling model\label{sec:expt}}
	One potential application of the proposed model is for inference of the interactions between suspended colloids through mapping onto the AHS model.  An important use case is the characterization of attractive interactions between macromolecular species such as proteins in concentrated solutions and under varying solution conditions.  Such characterization is needed for the purification, dewatering, and storage of biologics in pharmaceutical applications.  Utilizing a hydrodynamic and thermodynamic model for rigid spherical colloids to describe macromolecules which are neither spherical nor rigid has a long tradition in macromolecular sciences stemming primarily from application of the Stokes-Einstein relation to estimate the hydrodynamic radius.  In the context of proteins, transport properties in these solutions are commonly investigated using dynamic light scattering.  In light scattering experiments the correlation of the scattering intensity with time, $ t $, is used to compute the so-called intermediate scattering function, $ F( q, t ) $, which depends on a scattering wave vector $ q $.  For short correlation times, the value of this function is a direct measurement of the static structure factor, $ S( q ) = F( q, 0 ) $, and its rate of change can be used to compute a wave vector dependent diffusivity, $ D( q ) = - q^{-2}(d/dt) \log( F(q, t ) / F(q,0) ) $ as $ t \rightarrow 0 $.  As $ q \rightarrow 0 $, $ D( q ) $ is just the collective diffusivity $ D_C $, and is related to the sedimentation coefficient, $ U / U_0 $, of the macromolecule by the simple relation: 
\begin{equation}
		D_C = \frac{D_0}{S(0)} \frac{U}{U_0}. \label{eqn:dofq}
	\end{equation}
Here, $D_{0}$ is the diffusion coefficient of the macromolecule at infinite dilution.  Therefore, dynamic light scattering measurements for small scattering wave vectors can be used to simultaneously measure the collective diffusivity and $ S( 0 ) $, which is linearly proportional to the isothermal compressibility of the macromolecular component of the solution.  

For dilute solutions of macromolecules, measurements of these two quantities is part of the standard suite of macromolecular characterizations.  In the dilute limit, $ S(0) = 1 + k_S v c $ and $ D_C = D_0 ( 1 + ( k_S - k_H ) v c) $.  Here, $ c $ is the molar concentration of macromolecules in solution and $ v $ is the molar volume of the macromolecule.  The dimensionless coefficient $ k_S $ is purely thermodynamic in origin and linearly proportional to the second virial coefficient.  The dimensionless coefficient $ k_H $ describes the linear variations in the sedimentation coefficient of the macromolecular solution.  For the sticky sphere dispersions described in the previous sections, $ v = 4 \pi a^3 N_A / 3  $, with $ N_A $ Avogadro's number, $ k_S = -8 + 2 / \tau$, $ k_H = 6.55 - 0.89 / \tau $.  From the slope of $ D_C $ and $ S(0 ) $, the quantities: $ v k_S $ and $ v k_H $, can be computed.  If the sticky sphere model is deemed applicable, then the molar volume and the stickiness parameter are computed instead.  The advantage of this latter approach is that $ v $ and $ \tau $ can be used to locate the macromolecular solution on the AHS phase diagram.

	\begin{figure*}[t]
		\includegraphics[width=0.95\textwidth]{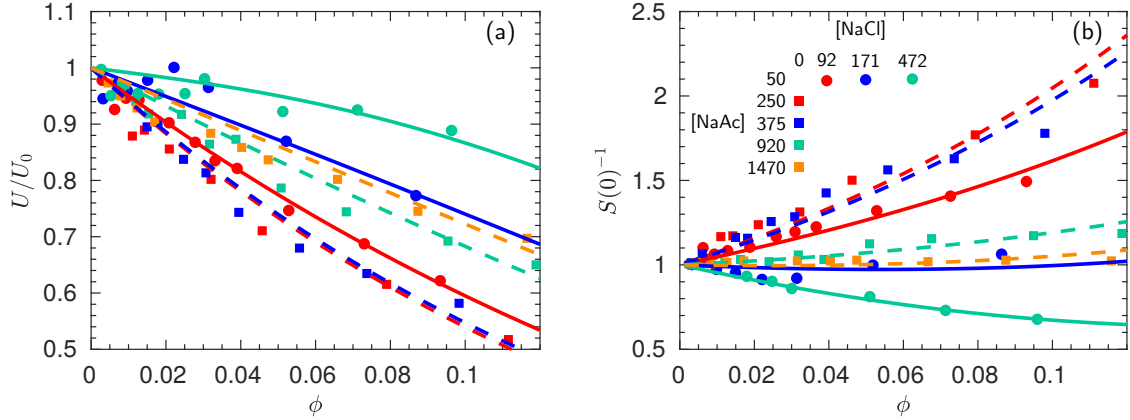}
		\caption{Sedimentation coefficient (a) and static structure factor (b) measured in Ref. \cite{muschol} for concentrated solutions of lysozyme.  Circles correspond to fixed buffer strength and varying added salt. Add concentrations are reported in units of mM.  Squares correspond to no added salt but varying buffer strength.  The solid and dashed lines are the best fits of the AHS models to the data for circles and squares, respectively.  The experimental data is presented as in Ref. \cite{muschol}, but with the concentration, $ c $, recast as the volume fraction on multiplication by the inferred molar volume.}
		\label{fig:lysozyme}
	\end{figure*}

Beyond the dilute limit, more sophisticated models are needed.  The AHS model for $ S( 0 ) $ is known analytically from the Percus-Yevick closure approximation applied to the Ornstein-Zernicke equation for the direct correlation function:
\begin{equation}
	S(0) = \left( \frac{\left( 1 - \phi \right)^{2}}{1 + 2\phi - \lambda\left(\phi,\tau\right) \, \phi \, \left( 1 - \phi \right)} \right)^{2},
\end{equation}
where,
\begin{align}
	\lambda\left( \phi, \tau \right) &= 6 \left( \frac{\tau}{\phi} + \frac{1}{1 - \phi} \right) - \\
		&\quad \left[ 36 \left( \frac{\tau}{\phi} + \frac{1}{1-\phi} \right)^{2} - 12 \frac{1 + 0.5\phi}{\phi \, \left( 1 - \phi \right)^{2}} \right]^{1/2}. \nonumber
\end{align}
From the present work, equation \eqref{eqn:settling_function} models the sedimentation coefficient, $ U / U_0 = D_C S(0) / D_0 $.  To relate experimental data to the two models, the colloidal volume fraction must be defined in terms of the molar concentration $ c $ as $ \phi = v c $.  Then, the experimental system has the properties of the sticky sphere solution at concentration $ c $ and parameterized by a molar volume $ v $ and stickiness parameter $ \tau $, just as in the dilute limit.  We envision three different modes of inference with these models:
\begin{itemize} 
	\item repeated measurements of $ S(0) $ and $ D_C $ at a particular concentration, $ c $ and in a particular solution condition (temperature, ionic strength, pH), denoted $ \mathbf{p} $, are used to determine the values of $ \tau(c, \mathbf{p}) $ and $ v(c, \mathbf{p}) $ for a corresponding AHS dispersion.  
	\item measurements of $ S(0) $ and $ D_C $ are made at different macromolecular concentrations and a particular solution condition, and a nonlinear least squares fit assuming a constant molar volume but concentration dependent stickiness parameter is used to find the set of parameters: $ v( \mathbf{p} ) $, $ \tau( c, \mathbf{p} ) $.
	\item $ S(0) $ and $ D_C $ are measured as function of the solution conditions at a particular macromolecule concentrations, and a nonlinear least squares fit of the data to the model is used to determine the parameters, $ v( \mathbf{p}) $ and $ \tau( \mathbf{p}) $ descriptive of the data across the entire range of concentrations.
\end{itemize}
Choice of inference scheme depends on the particular macromolecule under study, one could proceed from one scheme to the next should the assumptions of concentration independent molar volume and stickiness prove sound.

\citeauthor{muschol} used light scattering experiments to measure $ S(0) $ and $ D_C $ of concentrated lysozyme solutions at high ionic strengths for which long-range electrostatic repulsions are screened and the macromolecules behave much like sticky spheres\cite{piazza_lysozyme}.  Because we neither designed nor performed the experiments, we choose the third inference problem which determines the molar volume and stickiness parameter as a function of the solution conditions alone.  Lysozyme is a globular protein with an ellipsoidal shape having major and minor semi-axes of $ a = 2.75 $ nm and $ b = c = 1.65 $ nm, and typical molecular weight of $ 14 $ kDa\cite{lysozyme_ellipse}.  Based on the dimensions of the effective ellipsoid, a molar volume associated with the molecule: $ v = 4 \pi a b c / ( 3 N_A ) = $ 52 mM$^{-1}$ is anticipated.  The lysozyme is suspended in a sodium acetate buffer whose concentration is varied between $ 50 $ and $ 1470 $ mM.  Additionally, for a 50 mM NaAc buffer, NaCl is added to the solution at concentrations up to $ 472 $ mM.  For each solution condition, $ S(0)^{-1} $ and $ U / U_0 $ were reported for lysozyme concentrations up to 70 mg/mL.  On the basis of the estimated molar volume, the maximum concentration is equivalent to a volume fraction of approximately 10\%.   

Figure \ref{fig:lysozyme} depicts these measurements as well as fits of the AHS model under each solution condition.  The models do a remarkable job of capturing the experimental data across a range of volume fractions.  In particular, equation \eqref{eqn:settling_function}, appears to describe the settling behavior of the proteins very well in conditions for which the dilute limit expressions of Batchelor would fail.  Table \ref{tab:lysozyme} reports the molar volume and the stickiness parameters inferred from fits to the experimental data.  For 50 mM buffer, the molar volume averages 60 mM$^{-1}$, while the stickiness parameter decreases with added salt.  Increasing the ionic strength of the solution decreases the Debye layer thickness associated with electrostatic repulsions that stabilize the proteins against aggregation.  On decreasing the Debye layer thickness, the macromolecular solutions appear more sticky.  With no added salt, but higher buffer concentration, the molar volume of the lysozyme is larger and averages 62 mM$^{-1}$.  With increasing added buffer, the proteins also appear to grow stickier.

\begin{table}
\begin{ruledtabular}
\begin{tabular}{ccc}
& $[$NaAc$]$ = 50 mM & \\
$[$NaCl$]$ & $ \tau $ & $ v $ \\
\hline
92 mM& 0.59 & 55 mM$^{-1}$ \\
171 mM& 0.22 & 65 mM$^{-1}$ \\
472 mM& 0.16 & 59 mM$^{-1}$ \\
\hline \hline
& $[$NaCl$]$ = 0 mM & \\
$[$NaAc$]$ &$ \tau $& $v$ \\
\hline
250 mM & 2.7 & 61 mM$^{-1}$ \\
375 mM& 1.8 & 55 mM$^{-1}$ \\
920 mM& 0.29 & 65 mM$^{-1}$ \\
1470 mM& 0.24 & 67 mM$^{-1}$
\end{tabular}
\end{ruledtabular}
\caption{Stickiness parameter and molar volume inferred from nonlinear least squares fit of AHS model to experimental measurements of $ S(0) $ and $ U/ U_0 $ in lysozyme solutions\cite{muschol}.\label{tab:lysozyme}}
\end{table}

The two lysozyme solutions with [NaCl] = 92 mM and [NaAc] = 250 mM have ionic strengths that differ by less that 5\%.  The same is true of the ([NaCl], [NaAc]) pairs (171, 375) mM and (472, 920) mM.  Consequently, the electrostatic interactions between the proteins are similar screened in each of these cases.  However, the stickiness parameters and molar volumes inferred from the AHS hard sphere models show no correspondence.  This same lack of correspondence is evident in the $ S(0) $ and $ U / U_0 $ data itself.  It is known that there are specific ion effects in concentrated electrolyte solutions that sensitively affect the interactions between charged colloids such as proteins\cite{bsk}.  These specific effects alter the structure factor and transport properties of the lysozyme solutions in ways that are difficult to predict with classical DLVO theory\cite{dlvo}.  However, inference of the molar volume and stickiness parameter from fitting the AHS model to DLS data is sufficient to characterize solution in a way that yields an accurate description of thermodynamic properties (isothermal compressibility) and transport properties (sedimentation coefficient).

\section{Conclusion}	
We present coarse-grained simulations of the sedimentation of sticky particle suspensions that are shown to be in quantitative agreement with published experimental data. We systematically studied the effect of inter-particle attraction strength and volume fraction on the mean sedimentation rate of the suspensions, observing a smooth transition with increasing attraction strength from hard sphere sedimentation to non-monotonic promoted settling due to particle aggregation. We developed a simple model for the sedimentation rate by modifying a Richardson-Zaki-like expression to match Batchelor's dilute limit prediction and including a function that extrapolates between hard sphere-like and aggregated states. The model quantitatively describes the simulation data up to the percolation phase boundary, and an example application of this model to the study of collective protein diffusion is illustrated. The mapping of sticky particle suspensions onto an effective AHS model is generic, and can be used to infer sedimentation and diffusion behavior across a range concentrations from light scattering measurements made at only a few concentrations, even in suspensions where the particles are not uniform or spherical, provided the attractions are sufficiently short-ranged. 
	

\acknowledgements
J. Swan and A. Fiore gratefully acknowledge funding from the MIT Energy Initiative Shell Seed Fund and NSF Career Award CBET-1554398.

%


\begin{thebibliography}{38}%
\makeatletter
\providecommand \@ifxundefined [1]{%
 \@ifx{#1\undefined}
}%
\providecommand \@ifnum [1]{%
 \ifnum #1\expandafter \@firstoftwo
 \else \expandafter \@secondoftwo
 \fi
}%
\providecommand \@ifx [1]{%
 \ifx #1\expandafter \@firstoftwo
 \else \expandafter \@secondoftwo
 \fi
}%
\providecommand \natexlab [1]{#1}%
\providecommand \enquote  [1]{``#1''}%
\providecommand \bibnamefont  [1]{#1}%
\providecommand \bibfnamefont [1]{#1}%
\providecommand \citenamefont [1]{#1}%
\providecommand \href@noop [0]{\@secondoftwo}%
\providecommand \href [0]{\begingroup \@sanitize@url \@href}%
\providecommand \@href[1]{\@@startlink{#1}\@@href}%
\providecommand \@@href[1]{\endgroup#1\@@endlink}%
\providecommand \@sanitize@url [0]{\catcode `\\12\catcode `\$12\catcode
  `\&12\catcode `\#12\catcode `\^12\catcode `\_12\catcode `\%12\relax}%
\providecommand \@@startlink[1]{}%
\providecommand \@@endlink[0]{}%
\providecommand \url  [0]{\begingroup\@sanitize@url \@url }%
\providecommand \@url [1]{\endgroup\@href {#1}{\urlprefix }}%
\providecommand \urlprefix  [0]{URL }%
\providecommand \Eprint [0]{\href }%
\providecommand \doibase [0]{http://dx.doi.org/}%
\providecommand \selectlanguage [0]{\@gobble}%
\providecommand \bibinfo  [0]{\@secondoftwo}%
\providecommand \bibfield  [0]{\@secondoftwo}%
\providecommand \translation [1]{[#1]}%
\providecommand \BibitemOpen [0]{}%
\providecommand \bibitemStop [0]{}%
\providecommand \bibitemNoStop [0]{.\EOS\space}%
\providecommand \EOS [0]{\spacefactor3000\relax}%
\providecommand \BibitemShut  [1]{\csname bibitem#1\endcsname}%
\let\auto@bib@innerbib\@empty
\bibitem [{\citenamefont {Lebowitz}\ \emph {et~al.}(2002)\citenamefont
  {Lebowitz}, \citenamefont {Lewis},\ and\ \citenamefont {Schuck}}]{auc}%
  \BibitemOpen
  \bibfield  {author} {\bibinfo {author} {\bibfnamefont {J.}~\bibnamefont
  {Lebowitz}}, \bibinfo {author} {\bibfnamefont {M.~S.}\ \bibnamefont {Lewis}},
  \ and\ \bibinfo {author} {\bibfnamefont {P.}~\bibnamefont {Schuck}},\
  }\href@noop {} {\bibfield  {journal} {\bibinfo  {journal} {Protein Science}\
  }\textbf {\bibinfo {volume} {11}},\ \bibinfo {pages} {2067} (\bibinfo {year}
  {2002})}\BibitemShut {NoStop}%
\bibitem [{\citenamefont {Lerche}(2002)}]{shelflife}%
  \BibitemOpen
  \bibfield  {author} {\bibinfo {author} {\bibfnamefont {D.}~\bibnamefont
  {Lerche}},\ }\href@noop {} {\bibfield  {journal} {\bibinfo  {journal}
  {Journal of Dispersion Science and Technology}\ }\textbf {\bibinfo {volume}
  {23}},\ \bibinfo {pages} {699} (\bibinfo {year} {2002})}\BibitemShut
  {NoStop}%
\bibitem [{\citenamefont {Rulkens}\ \emph {et~al.}(1998)\citenamefont
  {Rulkens}, \citenamefont {Tichy},\ and\ \citenamefont
  {Grotenhuis}}]{pollution}%
  \BibitemOpen
  \bibfield  {author} {\bibinfo {author} {\bibfnamefont {W.}~\bibnamefont
  {Rulkens}}, \bibinfo {author} {\bibfnamefont {R.}~\bibnamefont {Tichy}}, \
  and\ \bibinfo {author} {\bibfnamefont {J.}~\bibnamefont {Grotenhuis}},\
  }\href@noop {} {\bibfield  {journal} {\bibinfo  {journal} {Water Science and
  Technology}\ }\textbf {\bibinfo {volume} {37}},\ \bibinfo {pages} {27}
  (\bibinfo {year} {1998})}\BibitemShut {NoStop}%
\bibitem [{\citenamefont {Partheniades}(1965)}]{erosion}%
  \BibitemOpen
  \bibfield  {author} {\bibinfo {author} {\bibfnamefont {E.}~\bibnamefont
  {Partheniades}},\ }\href@noop {} {\bibfield  {journal} {\bibinfo  {journal}
  {Journal of the Hydraulics Division}\ }\textbf {\bibinfo {volume} {91}},\
  \bibinfo {pages} {105} (\bibinfo {year} {1965})}\BibitemShut {NoStop}%
\bibitem [{\citenamefont {Roa}\ \emph {et~al.}(2015)\citenamefont {Roa},
  \citenamefont {Zholkovskiy},\ and\ \citenamefont
  {N{\"a}gele}}]{ultrafiltration}%
  \BibitemOpen
  \bibfield  {author} {\bibinfo {author} {\bibfnamefont {R.}~\bibnamefont
  {Roa}}, \bibinfo {author} {\bibfnamefont {E.~K.}\ \bibnamefont
  {Zholkovskiy}}, \ and\ \bibinfo {author} {\bibfnamefont {G.}~\bibnamefont
  {N{\"a}gele}},\ }\href@noop {} {\bibfield  {journal} {\bibinfo  {journal}
  {Soft matter}\ }\textbf {\bibinfo {volume} {11}},\ \bibinfo {pages} {4106}
  (\bibinfo {year} {2015})}\BibitemShut {NoStop}%
\bibitem [{\citenamefont {Guazzelli}\ and\ \citenamefont
  {Hinch}(2011)}]{guazzelli}%
  \BibitemOpen
  \bibfield  {author} {\bibinfo {author} {\bibfnamefont {E.}~\bibnamefont
  {Guazzelli}}\ and\ \bibinfo {author} {\bibfnamefont {J.}~\bibnamefont
  {Hinch}},\ }\href@noop {} {\bibfield  {journal} {\bibinfo  {journal} {Annual
  review of fluid mechanics}\ }\textbf {\bibinfo {volume} {43}},\ \bibinfo
  {pages} {97} (\bibinfo {year} {2011})}\BibitemShut {NoStop}%
\bibitem [{\citenamefont {Batchelor}(1972)}]{batchelor1}%
  \BibitemOpen
  \bibfield  {author} {\bibinfo {author} {\bibfnamefont {G.}~\bibnamefont
  {Batchelor}},\ }\href@noop {} {\bibfield  {journal} {\bibinfo  {journal}
  {Journal of Fluid Mechanics}\ }\textbf {\bibinfo {volume} {52}},\ \bibinfo
  {pages} {245} (\bibinfo {year} {1972})}\BibitemShut {NoStop}%
\bibitem [{\citenamefont {Batchelor}(1982)}]{batchelor2}%
  \BibitemOpen
  \bibfield  {author} {\bibinfo {author} {\bibfnamefont {G.}~\bibnamefont
  {Batchelor}},\ }\href@noop {} {\bibfield  {journal} {\bibinfo  {journal}
  {Journal of Fluid Mechanics}\ }\textbf {\bibinfo {volume} {119}},\ \bibinfo
  {pages} {379} (\bibinfo {year} {1982})}\BibitemShut {NoStop}%
\bibitem [{\citenamefont {Moncho-Jord{\'a}}\ \emph {et~al.}(2010)\citenamefont
  {Moncho-Jord{\'a}}, \citenamefont {Louis},\ and\ \citenamefont
  {Padding}}]{monchojorda1}%
  \BibitemOpen
  \bibfield  {author} {\bibinfo {author} {\bibfnamefont {A.}~\bibnamefont
  {Moncho-Jord{\'a}}}, \bibinfo {author} {\bibfnamefont {A.}~\bibnamefont
  {Louis}}, \ and\ \bibinfo {author} {\bibfnamefont {J.}~\bibnamefont
  {Padding}},\ }\href@noop {} {\bibfield  {journal} {\bibinfo  {journal}
  {Physical Review Letters}\ }\textbf {\bibinfo {volume} {104}},\ \bibinfo
  {pages} {068301} (\bibinfo {year} {2010})}\BibitemShut {NoStop}%
\bibitem [{\citenamefont {Lattuada}\ \emph {et~al.}(2016)\citenamefont
  {Lattuada}, \citenamefont {Buzzaccaro},\ and\ \citenamefont
  {Piazza}}]{lattuada}%
  \BibitemOpen
  \bibfield  {author} {\bibinfo {author} {\bibfnamefont {E.}~\bibnamefont
  {Lattuada}}, \bibinfo {author} {\bibfnamefont {S.}~\bibnamefont
  {Buzzaccaro}}, \ and\ \bibinfo {author} {\bibfnamefont {R.}~\bibnamefont
  {Piazza}},\ }\href@noop {} {\bibfield  {journal} {\bibinfo  {journal}
  {Physical Review Letters}\ }\textbf {\bibinfo {volume} {116}},\ \bibinfo
  {pages} {038301} (\bibinfo {year} {2016})}\BibitemShut {NoStop}%
\bibitem [{\citenamefont {Swan}\ and\ \citenamefont {Wang}(2016)}]{gang}%
  \BibitemOpen
  \bibfield  {author} {\bibinfo {author} {\bibfnamefont {J.~W.}\ \bibnamefont
  {Swan}}\ and\ \bibinfo {author} {\bibfnamefont {G.}~\bibnamefont {Wang}},\
  }\href {\doibase 10.1063/1.4939581} {\bibfield  {journal} {\bibinfo
  {journal} {Physics of Fluids}\ }\textbf {\bibinfo {volume} {28}},\ \bibinfo
  {pages} {011902} (\bibinfo {year} {2016})}\BibitemShut {NoStop}%
\bibitem [{\citenamefont {Usabiaga}\ \emph {et~al.}(2016)\citenamefont
  {Usabiaga}, \citenamefont {Kallemov}, \citenamefont {Delmotte}, \citenamefont
  {Bhalla}, \citenamefont {Griffith}, \citenamefont {Donev} \emph
  {et~al.}}]{multiblob}%
  \BibitemOpen
  \bibfield  {author} {\bibinfo {author} {\bibfnamefont {F.~B.}\ \bibnamefont
  {Usabiaga}}, \bibinfo {author} {\bibfnamefont {B.}~\bibnamefont {Kallemov}},
  \bibinfo {author} {\bibfnamefont {B.}~\bibnamefont {Delmotte}}, \bibinfo
  {author} {\bibfnamefont {A.~P.~S.}\ \bibnamefont {Bhalla}}, \bibinfo {author}
  {\bibfnamefont {B.~E.}\ \bibnamefont {Griffith}}, \bibinfo {author}
  {\bibfnamefont {A.}~\bibnamefont {Donev}},  \emph {et~al.},\ }\href@noop {}
  {\bibfield  {journal} {\bibinfo  {journal} {Communications in Applied
  Mathematics and Computational Science}\ }\textbf {\bibinfo {volume} {11}},\
  \bibinfo {pages} {217} (\bibinfo {year} {2016})}\BibitemShut {NoStop}%
\bibitem [{\citenamefont {Fiore}\ \emph {et~al.}(2017)\citenamefont {Fiore},
  \citenamefont {Balboa~Usabiaga}, \citenamefont {Donev},\ and\ \citenamefont
  {Swan}}]{pse}%
  \BibitemOpen
  \bibfield  {author} {\bibinfo {author} {\bibfnamefont {A.~M.}\ \bibnamefont
  {Fiore}}, \bibinfo {author} {\bibfnamefont {F.}~\bibnamefont
  {Balboa~Usabiaga}}, \bibinfo {author} {\bibfnamefont {A.}~\bibnamefont
  {Donev}}, \ and\ \bibinfo {author} {\bibfnamefont {J.~W.}\ \bibnamefont
  {Swan}},\ }\href@noop {} {\bibfield  {journal} {\bibinfo  {journal} {The
  Journal of Chemical Physics}\ }\textbf {\bibinfo {volume} {146}},\ \bibinfo
  {pages} {124116} (\bibinfo {year} {2017})}\BibitemShut {NoStop}%
\bibitem [{\citenamefont {Fiore}\ and\ \citenamefont
  {Swan}(2018)}]{pse-stresslet}%
  \BibitemOpen
  \bibfield  {author} {\bibinfo {author} {\bibfnamefont {A.~M.}\ \bibnamefont
  {Fiore}}\ and\ \bibinfo {author} {\bibfnamefont {J.~W.}\ \bibnamefont
  {Swan}},\ }\href@noop {} {\bibfield  {journal} {\bibinfo  {journal} {The
  Journal of chemical physics}\ }\textbf {\bibinfo {volume} {148}},\ \bibinfo
  {pages} {044114} (\bibinfo {year} {2018})}\BibitemShut {NoStop}%
\bibitem [{\citenamefont {Rotne}\ and\ \citenamefont {Prager}(1969)}]{rpy}%
  \BibitemOpen
  \bibfield  {author} {\bibinfo {author} {\bibfnamefont {J.}~\bibnamefont
  {Rotne}}\ and\ \bibinfo {author} {\bibfnamefont {S.}~\bibnamefont {Prager}},\
  }\href {\doibase http://dx.doi.org/10.1063/1.1670977} {\bibfield  {journal}
  {\bibinfo  {journal} {The Journal of Chemical Physics}\ }\textbf {\bibinfo
  {volume} {50}},\ \bibinfo {pages} {4831} (\bibinfo {year}
  {1969})}\BibitemShut {NoStop}%
\bibitem [{\citenamefont {Lindbo}\ and\ \citenamefont
  {Tornberg}(2010)}]{lindbo-tornberg}%
  \BibitemOpen
  \bibfield  {author} {\bibinfo {author} {\bibfnamefont {D.}~\bibnamefont
  {Lindbo}}\ and\ \bibinfo {author} {\bibfnamefont {A.-K.}\ \bibnamefont
  {Tornberg}},\ }\href@noop {} {\bibfield  {journal} {\bibinfo  {journal}
  {Journal of Computational Physics}\ }\textbf {\bibinfo {volume} {229}},\
  \bibinfo {pages} {8994} (\bibinfo {year} {2010})}\BibitemShut {NoStop}%
\bibitem [{\citenamefont {Keller}\ \emph {et~al.}(2000)\citenamefont {Keller},
  \citenamefont {Gould},\ and\ \citenamefont {Wathen}}]{keller}%
  \BibitemOpen
  \bibfield  {author} {\bibinfo {author} {\bibfnamefont {C.}~\bibnamefont
  {Keller}}, \bibinfo {author} {\bibfnamefont {N.~I.}\ \bibnamefont {Gould}}, \
  and\ \bibinfo {author} {\bibfnamefont {A.~J.}\ \bibnamefont {Wathen}},\
  }\href@noop {} {\bibfield  {journal} {\bibinfo  {journal} {SIAM Journal on
  Matrix Analysis and Applications}\ }\textbf {\bibinfo {volume} {21}},\
  \bibinfo {pages} {1300} (\bibinfo {year} {2000})}\BibitemShut {NoStop}%
\bibitem [{\citenamefont {Zick}\ and\ \citenamefont {Homsy}(1982)}]{zickhomsy}%
  \BibitemOpen
  \bibfield  {author} {\bibinfo {author} {\bibfnamefont {A.}~\bibnamefont
  {Zick}}\ and\ \bibinfo {author} {\bibfnamefont {G.}~\bibnamefont {Homsy}},\
  }\href@noop {} {\bibfield  {journal} {\bibinfo  {journal} {Journal of fluid
  mechanics}\ }\textbf {\bibinfo {volume} {115}},\ \bibinfo {pages} {13}
  (\bibinfo {year} {1982})}\BibitemShut {NoStop}%
\bibitem [{\citenamefont {Segre}\ \emph {et~al.}(1997)\citenamefont {Segre},
  \citenamefont {Herbolzheimer},\ and\ \citenamefont {Chaikin}}]{segre}%
  \BibitemOpen
  \bibfield  {author} {\bibinfo {author} {\bibfnamefont {P.}~\bibnamefont
  {Segre}}, \bibinfo {author} {\bibfnamefont {E.}~\bibnamefont
  {Herbolzheimer}}, \ and\ \bibinfo {author} {\bibfnamefont {P.}~\bibnamefont
  {Chaikin}},\ }\href@noop {} {\bibfield  {journal} {\bibinfo  {journal}
  {Physical Review Letters}\ }\textbf {\bibinfo {volume} {79}},\ \bibinfo
  {pages} {2574} (\bibinfo {year} {1997})}\BibitemShut {NoStop}%
\bibitem [{\citenamefont {Asakura}\ and\ \citenamefont
  {Oosawa}(1954)}]{ao_potential}%
  \BibitemOpen
  \bibfield  {author} {\bibinfo {author} {\bibfnamefont {S.}~\bibnamefont
  {Asakura}}\ and\ \bibinfo {author} {\bibfnamefont {F.}~\bibnamefont
  {Oosawa}},\ }\href@noop {} {\bibfield  {journal} {\bibinfo  {journal} {The
  Journal of Chemical Physics}\ }\textbf {\bibinfo {volume} {22}},\ \bibinfo
  {pages} {1255} (\bibinfo {year} {1954})}\BibitemShut {NoStop}%
\bibitem [{\citenamefont {Ladd}\ \emph {et~al.}(1995)\citenamefont {Ladd},
  \citenamefont {Gang}, \citenamefont {Zhu},\ and\ \citenamefont
  {Weitz}}]{laddweitz}%
  \BibitemOpen
  \bibfield  {author} {\bibinfo {author} {\bibfnamefont {A.~J.}\ \bibnamefont
  {Ladd}}, \bibinfo {author} {\bibfnamefont {H.}~\bibnamefont {Gang}}, \bibinfo
  {author} {\bibfnamefont {J.}~\bibnamefont {Zhu}}, \ and\ \bibinfo {author}
  {\bibfnamefont {D.}~\bibnamefont {Weitz}},\ }\href@noop {} {\bibfield
  {journal} {\bibinfo  {journal} {Physical Review E}\ }\textbf {\bibinfo
  {volume} {52}},\ \bibinfo {pages} {6550} (\bibinfo {year}
  {1995})}\BibitemShut {NoStop}%
\bibitem [{\citenamefont {Noro}\ and\ \citenamefont {Frenkel}(2000)}]{noro}%
  \BibitemOpen
  \bibfield  {author} {\bibinfo {author} {\bibfnamefont {M.~G.}\ \bibnamefont
  {Noro}}\ and\ \bibinfo {author} {\bibfnamefont {D.}~\bibnamefont {Frenkel}},\
  }\href@noop {} {\bibfield  {journal} {\bibinfo  {journal} {The Journal of
  Chemical Physics}\ }\textbf {\bibinfo {volume} {113}},\ \bibinfo {pages}
  {2941} (\bibinfo {year} {2000})}\BibitemShut {NoStop}%
\bibitem [{\citenamefont {Cichocki}\ and\ \citenamefont
  {Sadlej}(2005)}]{cichocki}%
  \BibitemOpen
  \bibfield  {author} {\bibinfo {author} {\bibfnamefont {B.}~\bibnamefont
  {Cichocki}}\ and\ \bibinfo {author} {\bibfnamefont {K.}~\bibnamefont
  {Sadlej}},\ }\href@noop {} {\bibfield  {journal} {\bibinfo  {journal} {EPL
  (Europhysics Letters)}\ }\textbf {\bibinfo {volume} {72}},\ \bibinfo {pages}
  {936} (\bibinfo {year} {2005})}\BibitemShut {NoStop}%
\bibitem [{\citenamefont {Baxter}(1968)}]{baxter}%
  \BibitemOpen
  \bibfield  {author} {\bibinfo {author} {\bibfnamefont {R.}~\bibnamefont
  {Baxter}},\ }\href@noop {} {\bibfield  {journal} {\bibinfo  {journal} {The
  Journal of Chemical Physics}\ }\textbf {\bibinfo {volume} {49}},\ \bibinfo
  {pages} {2770} (\bibinfo {year} {1968})}\BibitemShut {NoStop}%
\bibitem [{\citenamefont {Percus}\ and\ \citenamefont
  {Yevick}(1958)}]{percus-yevick}%
  \BibitemOpen
  \bibfield  {author} {\bibinfo {author} {\bibfnamefont {J.~K.}\ \bibnamefont
  {Percus}}\ and\ \bibinfo {author} {\bibfnamefont {G.~J.}\ \bibnamefont
  {Yevick}},\ }\href@noop {} {\bibfield  {journal} {\bibinfo  {journal}
  {Physical Review}\ }\textbf {\bibinfo {volume} {110}},\ \bibinfo {pages} {1}
  (\bibinfo {year} {1958})}\BibitemShut {NoStop}%
\bibitem [{\citenamefont {Regnaut}\ and\ \citenamefont
  {Ravey}(1990)}]{regnaut2}%
  \BibitemOpen
  \bibfield  {author} {\bibinfo {author} {\bibfnamefont {C.}~\bibnamefont
  {Regnaut}}\ and\ \bibinfo {author} {\bibfnamefont {J.}~\bibnamefont
  {Ravey}},\ }\href@noop {} {\bibfield  {journal} {\bibinfo  {journal} {The
  Journal of Chemical Physics}\ }\textbf {\bibinfo {volume} {92}},\ \bibinfo
  {pages} {3250} (\bibinfo {year} {1990})}\BibitemShut {NoStop}%
\bibitem [{\citenamefont {Richardson}\ and\ \citenamefont
  {Zaki}(1954)}]{richardson_zaki}%
  \BibitemOpen
  \bibfield  {author} {\bibinfo {author} {\bibfnamefont {J.}~\bibnamefont
  {Richardson}}\ and\ \bibinfo {author} {\bibfnamefont {W.}~\bibnamefont
  {Zaki}},\ }\href@noop {} {\bibfield  {journal} {\bibinfo  {journal} {Chemical
  Engineering Science}\ }\textbf {\bibinfo {volume} {3}},\ \bibinfo {pages}
  {65} (\bibinfo {year} {1954})}\BibitemShut {NoStop}%
\bibitem [{\citenamefont {Xue}\ \emph {et~al.}(1992)\citenamefont {Xue},
  \citenamefont {Herbolzheimer}, \citenamefont {Rutgers}, \citenamefont
  {Russel},\ and\ \citenamefont {Chaikin}}]{piazza_squares}%
  \BibitemOpen
  \bibfield  {author} {\bibinfo {author} {\bibfnamefont {J.-Z.}\ \bibnamefont
  {Xue}}, \bibinfo {author} {\bibfnamefont {E.}~\bibnamefont {Herbolzheimer}},
  \bibinfo {author} {\bibfnamefont {M.}~\bibnamefont {Rutgers}}, \bibinfo
  {author} {\bibfnamefont {W.}~\bibnamefont {Russel}}, \ and\ \bibinfo {author}
  {\bibfnamefont {P.}~\bibnamefont {Chaikin}},\ }\href@noop {} {\bibfield
  {journal} {\bibinfo  {journal} {Physical review letters}\ }\textbf {\bibinfo
  {volume} {69}},\ \bibinfo {pages} {1715} (\bibinfo {year}
  {1992})}\BibitemShut {NoStop}%
\bibitem [{\citenamefont {Bacri}\ \emph {et~al.}(1986)\citenamefont {Bacri},
  \citenamefont {Frenois}, \citenamefont {Hoyos}, \citenamefont {Perzynski},
  \citenamefont {Rakotomalala},\ and\ \citenamefont
  {Salin}}]{piazza_triangles}%
  \BibitemOpen
  \bibfield  {author} {\bibinfo {author} {\bibfnamefont {J.-C.}\ \bibnamefont
  {Bacri}}, \bibinfo {author} {\bibfnamefont {C.}~\bibnamefont {Frenois}},
  \bibinfo {author} {\bibfnamefont {M.}~\bibnamefont {Hoyos}}, \bibinfo
  {author} {\bibfnamefont {R.}~\bibnamefont {Perzynski}}, \bibinfo {author}
  {\bibfnamefont {N.}~\bibnamefont {Rakotomalala}}, \ and\ \bibinfo {author}
  {\bibfnamefont {D.}~\bibnamefont {Salin}},\ }\href@noop {} {\bibfield
  {journal} {\bibinfo  {journal} {EPL (Europhysics Letters)}\ }\textbf
  {\bibinfo {volume} {2}},\ \bibinfo {pages} {123} (\bibinfo {year}
  {1986})}\BibitemShut {NoStop}%
\bibitem [{\citenamefont {Buzzaccaro}\ \emph {et~al.}(2008)\citenamefont
  {Buzzaccaro}, \citenamefont {Tripodi}, \citenamefont {Rusconi}, \citenamefont
  {Vigolo},\ and\ \citenamefont {Piazza}}]{piazza_utriangles}%
  \BibitemOpen
  \bibfield  {author} {\bibinfo {author} {\bibfnamefont {S.}~\bibnamefont
  {Buzzaccaro}}, \bibinfo {author} {\bibfnamefont {A.}~\bibnamefont {Tripodi}},
  \bibinfo {author} {\bibfnamefont {R.}~\bibnamefont {Rusconi}}, \bibinfo
  {author} {\bibfnamefont {D.}~\bibnamefont {Vigolo}}, \ and\ \bibinfo {author}
  {\bibfnamefont {R.}~\bibnamefont {Piazza}},\ }\href@noop {} {\bibfield
  {journal} {\bibinfo  {journal} {Journal of Physics: Condensed Matter}\
  }\textbf {\bibinfo {volume} {20}},\ \bibinfo {pages} {494219} (\bibinfo
  {year} {2008})}\BibitemShut {NoStop}%
\bibitem [{\citenamefont {De~Kruif}\ \emph {et~al.}(1987)\citenamefont
  {De~Kruif}, \citenamefont {Jansen},\ and\ \citenamefont
  {Vrij}}]{piazza_diamonds}%
  \BibitemOpen
  \bibfield  {author} {\bibinfo {author} {\bibfnamefont {C.}~\bibnamefont
  {De~Kruif}}, \bibinfo {author} {\bibfnamefont {J.}~\bibnamefont {Jansen}}, \
  and\ \bibinfo {author} {\bibfnamefont {A.}~\bibnamefont {Vrij}},\ }\href@noop
  {} {\bibfield  {journal} {\bibinfo  {journal} {Physics of Complex and
  Supramolecular Fluids}\ ,\ \bibinfo {pages} {315}} (\bibinfo {year}
  {1987})}\BibitemShut {NoStop}%
\bibitem [{\citenamefont {Buscall}\ \emph {et~al.}(1982)\citenamefont
  {Buscall}, \citenamefont {Goodwin}, \citenamefont {Ottewill},\ and\
  \citenamefont {Tadros}}]{piazza_ccircles}%
  \BibitemOpen
  \bibfield  {author} {\bibinfo {author} {\bibfnamefont {R.}~\bibnamefont
  {Buscall}}, \bibinfo {author} {\bibfnamefont {J.}~\bibnamefont {Goodwin}},
  \bibinfo {author} {\bibfnamefont {R.}~\bibnamefont {Ottewill}}, \ and\
  \bibinfo {author} {\bibfnamefont {T.~F.}\ \bibnamefont {Tadros}},\
  }\href@noop {} {\bibfield  {journal} {\bibinfo  {journal} {Journal of Colloid
  and Interface Science}\ }\textbf {\bibinfo {volume} {85}},\ \bibinfo {pages}
  {78} (\bibinfo {year} {1982})}\BibitemShut {NoStop}%
\bibitem [{\citenamefont {Ladd}(1990)}]{piazza_ocircles}%
  \BibitemOpen
  \bibfield  {author} {\bibinfo {author} {\bibfnamefont {A.~J.}\ \bibnamefont
  {Ladd}},\ }\href@noop {} {\bibfield  {journal} {\bibinfo  {journal} {The
  Journal of chemical physics}\ }\textbf {\bibinfo {volume} {93}},\ \bibinfo
  {pages} {3484} (\bibinfo {year} {1990})}\BibitemShut {NoStop}%
\bibitem [{\citenamefont {Muschol}\ and\ \citenamefont
  {Rosenberger}(1995)}]{muschol}%
  \BibitemOpen
  \bibfield  {author} {\bibinfo {author} {\bibfnamefont {M.}~\bibnamefont
  {Muschol}}\ and\ \bibinfo {author} {\bibfnamefont {F.}~\bibnamefont
  {Rosenberger}},\ }\href {\doibase 10.1063/1.469891} {\bibfield  {journal}
  {\bibinfo  {journal} {The Journal of Chemical Physics}\ }\textbf {\bibinfo
  {volume} {103}},\ \bibinfo {pages} {10424} (\bibinfo {year} {1995})},\
  \Eprint {http://arxiv.org/abs/https://doi.org/10.1063/1.469891}
  {https://doi.org/10.1063/1.469891} \BibitemShut {NoStop}%
\bibitem [{\citenamefont {Piazza}\ \emph {et~al.}(1998)\citenamefont {Piazza},
  \citenamefont {Peyre},\ and\ \citenamefont {Degiorgio}}]{piazza_lysozyme}%
  \BibitemOpen
  \bibfield  {author} {\bibinfo {author} {\bibfnamefont {R.}~\bibnamefont
  {Piazza}}, \bibinfo {author} {\bibfnamefont {V.}~\bibnamefont {Peyre}}, \
  and\ \bibinfo {author} {\bibfnamefont {V.}~\bibnamefont {Degiorgio}},\ }\href
  {\doibase 10.1103/PhysRevE.58.R2733} {\bibfield  {journal} {\bibinfo
  {journal} {Phys. Rev. E}\ }\textbf {\bibinfo {volume} {58}},\ \bibinfo
  {pages} {R2733} (\bibinfo {year} {1998})}\BibitemShut {NoStop}%
\bibitem [{\citenamefont {Dubin}\ \emph {et~al.}(1971)\citenamefont {Dubin},
  \citenamefont {Clark},\ and\ \citenamefont {Benedek}}]{lysozyme_ellipse}%
  \BibitemOpen
  \bibfield  {author} {\bibinfo {author} {\bibfnamefont {S.~B.}\ \bibnamefont
  {Dubin}}, \bibinfo {author} {\bibfnamefont {N.~A.}\ \bibnamefont {Clark}}, \
  and\ \bibinfo {author} {\bibfnamefont {G.~B.}\ \bibnamefont {Benedek}},\
  }\href {\doibase 10.1063/1.1674810} {\bibfield  {journal} {\bibinfo
  {journal} {The Journal of Chemical Physics}\ }\textbf {\bibinfo {volume}
  {54}},\ \bibinfo {pages} {5158} (\bibinfo {year} {1971})},\ \Eprint
  {http://arxiv.org/abs/https://doi.org/10.1063/1.1674810}
  {https://doi.org/10.1063/1.1674810} \BibitemShut {NoStop}%
\bibitem [{\citenamefont {Bazant}\ \emph {et~al.}(2011)\citenamefont {Bazant},
  \citenamefont {Storey},\ and\ \citenamefont {Kornyshev}}]{bsk}%
  \BibitemOpen
  \bibfield  {author} {\bibinfo {author} {\bibfnamefont {M.~Z.}\ \bibnamefont
  {Bazant}}, \bibinfo {author} {\bibfnamefont {B.~D.}\ \bibnamefont {Storey}},
  \ and\ \bibinfo {author} {\bibfnamefont {A.~A.}\ \bibnamefont {Kornyshev}},\
  }\href {\doibase 10.1103/PhysRevLett.106.046102} {\bibfield  {journal}
  {\bibinfo  {journal} {Phys. Rev. Lett.}\ }\textbf {\bibinfo {volume} {106}},\
  \bibinfo {pages} {046102} (\bibinfo {year} {2011})}\BibitemShut {NoStop}%
\bibitem [{\citenamefont {Bostr\"om}\ \emph {et~al.}(2001)\citenamefont
  {Bostr\"om}, \citenamefont {Williams},\ and\ \citenamefont {Ninham}}]{dlvo}%
  \BibitemOpen
  \bibfield  {author} {\bibinfo {author} {\bibfnamefont {M.}~\bibnamefont
  {Bostr\"om}}, \bibinfo {author} {\bibfnamefont {D.~R.~M.}\ \bibnamefont
  {Williams}}, \ and\ \bibinfo {author} {\bibfnamefont {B.~W.}\ \bibnamefont
  {Ninham}},\ }\href {\doibase 10.1103/PhysRevLett.87.168103} {\bibfield
  {journal} {\bibinfo  {journal} {Phys. Rev. Lett.}\ }\textbf {\bibinfo
  {volume} {87}},\ \bibinfo {pages} {168103} (\bibinfo {year}
  {2001})}\BibitemShut {NoStop}%
\end{thebibliography}
\end{document}